\documentclass[floatfix,twocolumn,amsmath,amssymb,pre,showpacs]{revtex4}

\usepackage{graphicx}
 \usepackage{placeins}
\usepackage{dcolumn}
\usepackage{bm}
\usepackage{amssymb}
\usepackage{amsmath}
\usepackage{amsthm,amsfonts}

\begin{document}
\bibliographystyle{aip}

\title{Non-trivial effect of dephasing: Enhancement of rectification of spin current in graded XX chains}

\author{ Saulo H. S. Silva$^{1}$, Gabriel T. Landi$^{2}$, and Emmanuel Pereira$^{1}$}
 \email{emmanuel@fisica.ufmg.br}
\affiliation{$^{1}$Departamento de F\'{\i}sica, Instituto de Ci\^encias Exatas, Universidade Federal de Minas Gerais, 30123-970, Belo Horizonte, Minas Gerais, Brazil\\
$^{2}$Instituto de Física da Universidade de São Paulo, 05314-970 São Paulo, Brazil\\
}

\begin{abstract}
In order to reveal  mechanisms to control and manipulate spin currents, we perform a detailed investigation of the dephasing effects in the open $XX$ model with a Lindblad dynamics involving global dissipators and thermal baths. Specifically, we consider dephasing noise modelled by current preserving Lindblad dissipators acting on graded versions of these spin systems, that is, systems in which the magnetic field and/or the spin interaction are growing (decreasing) along the chain. In our analysis, we study the non-equilibrium steady-state via the covariance matrix using the Jordan-Wigner approach  to compute the spin currents. We find that the interplay between dephasing and graded systems gives rise to a non trivial behavior: when we have homogeneous magnetic field and graded interactions we have rectification enhancement mechanims, and when we have fully graded system we can control the spin current in order to keep the direction of the particle/spin flow even with inverted baths. We describe our result in detailed numerical analisys and we see that rectification in this simple model indicates that the phenomenon may be of general occurrence in quantum spin systems.

\end{abstract}


\def \Z {\mathbb{Z}}
\def \R {\mathbb{R}}
\def \La {\Lambda}
\def \la {\lambda}
\def \ck {l}
\def \F {\mathcal{F}}
\def \M {\mathcal{M}}
\newcommand {\md} [1] {\mid\!#1\!\mid}
\newcommand {\be} {\begin{equation}}
\newcommand {\ee} {\end{equation}}
\newcommand {\ben} {\begin{equation*}}
\newcommand {\een} {\end{equation*}}
\newcommand {\bg} {\begin{gather}}
\newcommand {\eg} {\end{gather}}
\newcommand {\ba} {\begin{align}}
\newcommand {\ea} {\end{align}}
\newcommand {\tit} [1] {``#1''}


\maketitle

\let\a=\alpha \let\b=\beta \let\d=\delta \let\e=\varepsilon
\let\f=\varphi \let\g=\gamma \let\h=\eta    \let\k=\kappa \let\l=\lambda
\let\m=\mu \let\n=\nu \let\o=\omega    \let\p=\pi \let\ph=\varphi
\let\r=\rho \let\s=\sigma \let\t=\tau \let\th=\vartheta
\let\y=\upsilon \let\x=\xi \let\z=\zeta
\let\D=\delta \let\G=\Gamma \let\L=\Lambda \let\Th=\Theta
\let\P=\Pi \let\Ps=\Psi \let\Si=\Sigma \let\X=\Xi
\let\Y=\Upsilon

\section{Introduction}

Transport in quantum devices has been receiving increasing attention due to the possibility of building smaller and smaller systems, which in turn has sharpened our understanding of non-classical effects on fluxes of energy or particles. The comprehension of these behaviors, precisely, the derivation of transport laws from the underlying microscopic dynamics, is one of the fundamental issues of nonequilibrium statistical physics. 

It also deserves attention the investigation of transport in low dimensional systems, which raises
interesting problems, both on classical and quantum regimes \cite{lepri,bertini,adhar,gtlandi}.

An important and recurrently studied transport property is the existence of rectification, namely, a preferential direction for the flow. Many works are devoted to study  what types of interactions the system has to present to guarantee rectification \cite{bli,BLiRMP,suff,PLA,BLL}.

In the quantum regime, open spin quantum systems governed by Lindblad equations have shown to 
present rectification and other interesting  behaviors with promising applications for the manipulation of the energy/spin flow \cite{barra,oneway,landi}. To give an example, we recall studies involving an Ising chain with thermal reservoirs attached to its ends: for a junction with two spins and a longitudinal field, it has been shown that perfect rectification occurs  \cite{optimal}.  Considering systems larger than two spins in the Ising chain, in Ref. \cite{perfect emmanuel}, conditions to keep the perfect rectification are presented. For more complex systems, such as the $XX$ chain, it is observed in \cite{sxx} that the rectification factor does not tend to zero at the thermodynamic limit.

In this context, an interesting problem  is the effect of dephasing noise in the quantum transport.
In particular, a problem to be considered in the present paper, it is of interest the effect of dephasing in the rectification property. 

The rule played by dephasing in the transport has been recurrently investigated, mainly in  \textit{boundary driven} systems.  As an example we cite interesting results described in Ref\cite{zn,asadian}: for any non-zero dephasing strengths, free tight-binding models typically become diffusive in the thermodynamic limit .  In Ref\cite{arthur}, it is demonstrated that the interplay between dephasing and  a particualr type of on-site potential, namelly quasi-periodic potentials, gives rise to an enhancement of transport, increasing the systems conductivity. In Ref\cite{mendoza} a detailed investigation of the heat flow on an $XXZ$ chain is performed when bulk dephasing takes place, both on the weakly-interacting and strongly interacting regimes.

It is important to recall that the $XXZ$ chains are the prototypes for open quantum spin systems.  In 
particular, the rectification phenomena has already been  investigated in these models. In Ref.\cite{GL1}, it is shown the absence of spin rectification in the system with zero anisotropy  parameter $\Delta$ (coefficient of $\sigma_{j}^{z}\sigma_{j+1}^{z}$) and for $\Delta \neq 0$, rectification is observed. 
Is also well known that the $XXZ$ model can be mapped into another problem: bosons with creation and annihilation operators, with quadratic terms and a quartic one proportional to $\Delta$ (Tonks-Girardeau model). The vanishing of rectification in the absence of the quartic term has an analogy with the case of classical oscillators, where it is known that there is no rectification in the absence of anharmonicity (or other effect beyond
pure harmonicity).

Motivated by the vast transport properties of open quantum systems and by the interesting effects of dephasing noise on transport, in the present work we perform an analytical and numerical detailed investigation of the $XX$ spin $1/2$ model, subject to thermal baths  and dephasing noise. Using the global Lindblad master equations, we focus  in a specific type of asymmetric systems, the \textit{graded} chains.  We remark that graded models have already demonstrated to be precise systems for the occurrence of rectification \cite{lucasgraded,graded,eplemmanuel}. In the present paper, we show a very interesting and non trivial effect: the addition of dephasing (which, we recall, means a kind of noise), in some cases, may increase the spin current and also the spin rectification in these graded $XX$ spin models.

The rest of the paper is organized as follows. In section II, we introduce
the model and some preliminary details about the study of the \textit{NESS}. In
section III, we describe the currents and some properties using the covariance matrix. In section IV, we perfom numerical results for the spin rectification  with dephasing. Section V is devoted to concluding remarks.


\section{Model and Preliminary Details}
Our model under study is the  one-dimensional quantum $XX$ spin chain with $N$ sites, described by the Hamiltonian 
\begin{equation}\label{eq:1}
    H=\sum_{j=1}^{N}\frac{h_{j}}{2}\sigma_{j}^{z}+\frac{1}{2}\sum_{j=1}^{N-1}\alpha_{j}(\sigma_{j}^{x}\sigma_{j+1}^{x}+\sigma_{j}^{y}\sigma_{j+1}^{y})\quad,
\end{equation}
\noindent
where  $\sigma_{j}^{i}$ are the usual Pauli matrices, $h_{j}$ is the external magnetic field acting on site $j$ and $\alpha_{j}$ is the exchange interaction between spins $j$ and $j+1$. The rectification and the fluxes present on the system will be directly associated with the asymmetry of the coefficients $h_{j}$ and $\alpha_{j}$ with respect to the left-right reflection of the chain.

The system is coupled on the first and last sites to thermal reservoirs, kept at temperatures $T_{L}$ and $T_{R}$, respectively. The thermal baths are modeled by an infinite number of bosonic degrees of freedom given by the following Hamiltonian

\begin{equation}\label{eq:2}
    H_{E}^{i}=\sum_{l}\Omega_{i,l}a_{i,l}^{\dagger}a_{i,l}\quad,
\end{equation}
\noindent
where $a_{i,l}$ are a set of independent bosonic operators and $\Omega_{i,l}$ are the corresponding frequencies, which we assume to take on a quasi-continuum of values in the interval $[0,\infty)$. The interaction with the first and last sites are assumed to take the form

\begin{equation}\label{eq:3}
\begin{aligned}
    H_{I}^{L}&=\sigma_{1}^{x}\sum_{i}g_{i}(a_{L,i}^{\dagger}+a_{L,i})\\
     H_{I}^{R}&=\sigma_{N}^{x}\sum_{i}g_{i}(a_{R,i}^{\dagger}+a_{R,i})\quad.
    \end{aligned}
\end{equation}

In order to proceed with the study of the currents, we recast the problem as a Lindblad master equation in the \textit{weak coupling regime} \cite{breuer}, where the time evolution of the system's density matrix $\rho$ is given by

\begin{equation}\label{eq:4}
    \frac{d\rho}{dt}=-i[H,\rho]+\mathcal{D}_{L}+\mathcal{D}_{R}\quad,
\end{equation}
\noindent
where $\mathcal{D}_{L}$ and $\mathcal{D}_{R}$ are the Lindblad dissipators associated to the baths. It is possible to derive them from Eq. \eqref{eq:3} using the method of eigenoperators \cite{breuer}.

To obtain a better representation of the Hamiltonian  we transform it in terms of  $\sigma_{l}^{+}$ and $\sigma_{l}^{-}$ operators given by
\begin{equation}\label{eq:5}
    \begin{aligned}
        \sigma_{l}^{+}&=\frac{1}{2}(\sigma_{l}^{x}+i\sigma_{l}^{y})\\
       \sigma_{l}^{-}&=\frac{1}{2}(\sigma_{l}^{x}-i\sigma_{l}^{y})\quad,\\
    \end{aligned}
\end{equation}
then the Hamiltonian in \eqref{eq:1} becomes 
\begin{equation}\label{eq:6}
    H=\sum_{j=1}^{N}\frac{h_{j}}{2}(\sigma_{j}^{+}\sigma_{j}^{-} -1/2)+\frac{1}{2}\sum_{j=1}^{N-1}\alpha_{j}(\sigma_{j}^{+}\sigma_{j+1}^{-}+\sigma_{j}^{-}\sigma_{j+1}^{+})\quad.
\end{equation}

In order to study the dissipators in the Lindblad equation we must diagonalize $H$. To do this we use a fermionic representation through the Jordan-Wigner transformation~\cite{L3,L4} given by:

\begin{equation}\label{eq:7}
\begin{aligned}
    &\eta_{l}=Q_{l}\sigma_{l}^{-}\\
    \\
    &\eta_{l}^{\dagger}=Q_{l}\sigma_{l}^{+}\quad,
    \end{aligned}
\end{equation}
\noindent
where $Q_{l}=\prod_{j=1}^{l-1}(-\sigma_{j}^{z})$.

Following this transformation, the Hamiltonian is given by a quadratic form in the fermionic operators
\begin{equation}\label{eq:8}
\begin{aligned}
    H&=\sum_{j=1}^{N}h_{j}\eta_{j}^{\dagger}\eta_{j}+\sum_{j=1}^{N-1}\alpha_{j}(\eta_{j}^{\dagger}\eta_{j+1}+\eta_{j+1}^{\dagger}\eta_{j})\\
    &=\sum_{n,m}W_{n,m}\eta_{n}^{\dagger}\eta_{m}\quad,
    \end{aligned}
\end{equation}
\noindent
where $W_{n,m}$ is a matrix with entries $W_{j,j}=h_{j}$ and $W_{j,j+1}=W_{j+1,j}=\alpha_{j}$.

It is possible  to put $H$ in diagonal form, that is, we first diagonalize the matrix $W$. Since it is symmetric, it may be diagonalized by an orthogonal transformation $S_{n,k}$ $(S^{\dagger}S=1)$ as
\begin{equation}\label{eq:9}
    W_{n,m}=\sum_{k=1}^{N}\epsilon_{k}S_{n,k}S_{m,k}\quad.
\end{equation}

Here we define a new set of fermionic operators

\begin{equation}\label{eq:10}
    \Tilde{\eta_{j}}=\sum_{k=1}^{N}S_{j,k}\eta_{k}\quad,
\end{equation}
\noindent
in terms of which Eq. \eqref{eq:9} becomes

\begin{equation}\label{eq:11}
    H=\sum_{k=1}^{N}\epsilon_{k}\Tilde{\eta_{k}}^{\dagger}\Tilde{\eta_{k}}\quad.
\end{equation}

As derived in a previous work\cite{sxx}, the dissipators are given in terms of this new set of fermionic operators:

\begin{equation}\label{eq:12}
\begin{aligned}
    \mathcal{D}_{L}(\rho)&=\sum_{k=1}^{N}\gamma (S_{1,k}^{-1})^{2}\chi_{L,k}\left\{[1-f_{L,k}]\left[\Tilde{\eta_{k}}\rho\Tilde{\eta_{k}}^{\dagger}-\frac{1}{2}\{\Tilde{\eta_{k}}^{\dagger}\Tilde{\eta_{k}},\rho\}\right]\right.\\
    &\left.+f_{L,k}\left[\Tilde{\eta_{k}}^{\dagger}\rho\Tilde{\eta_{k}}-\frac{1}{2}\{\Tilde{\eta_{k}}\Tilde{\eta_{k}}^{\dagger},\rho\}\right]\right\}\quad.
    \end{aligned}
\end{equation}

And for the site coupled to the right reservoir

\begin{equation}\label{eq:13}
\begin{aligned}
    \mathcal{D}_{R}(\rho)=\sum_{k=1}^{N}\gamma &(S_{N,k}^{-1})^{2}\chi_{R,k}\left\{[1-f_{R,k}]\times\right.\\
    &\left[\Tilde{\eta_{k}}e^{i\pi\mathcal{N}}\rho e^{i\pi\mathcal{N}}\Tilde{\eta_{k}}^{\dagger}-\frac{1}{2}\{\Tilde{\eta_{k}}^{\dagger}\Tilde{\eta_{k}},\rho\}\right]\\
    &\left.+f_{R,k}\left[\Tilde{\eta_{k}}^{\dagger}e^{i\pi\mathcal{N}}\rho e^{i\pi\mathcal{N}}\Tilde{\eta_{k}}-\frac{1}{2}\{\Tilde{\eta_{k}}\Tilde{\eta_{k}}^{\dagger},\rho\}\right]\right\}\quad.
    \end{aligned}
\end{equation}
where $\chi_{k,L(R)}$ is a temperature-dependent function, $f_{k,L(R)}$ is the Fermi-Dirac distribution,

\begin{equation}\label{eq:14}
    \begin{aligned}
           f_{L(R),k}&=\frac{1}{e^{\epsilon_{k}}/T_{L(R)}+1}\\
           \\
        \chi_{L(R),k}&=\coth\left(\frac{|\epsilon_{k}|}{2T_{L(R)}}\right)\quad,
    \end{aligned}
\end{equation}
and $\mathcal{N}$ is the total number of fermions:

\begin{equation*}
    \mathcal{N}=\sum_{i=1}^{N}\Tilde{\eta}_{i}^{\dagger}\Tilde{\eta}_{i}.
\end{equation*}

The expression for the dissipators can be writen in a more elegant form given by

\begin{equation*}
    \begin{aligned}
        D_{k}^{L}=A_{L}^{k}&\left[\Tilde{\eta}_{k}\rho\Tilde{\eta}_{k}^{\dagger}-\frac{1}{2}\left\{\Tilde{\eta}_{k}^{\dagger}\Tilde{\eta},\rho\right\}\right]\\
        \\
        &+B_{L}^{k}\left[\Tilde{\eta}_{k}^{\dagger}\rho\Tilde{\eta}_{k}-\frac{1}{2}\left\{\Tilde{\eta}_{k}\Tilde{\eta}_{k}^{\dagger},\rho\right\}\right]
    \end{aligned}
\end{equation*}
\begin{equation*}
    \begin{aligned}
        D_{k}^{R}=A_{R}^{k}&\left[\Tilde{\eta}_{k}e^{i\pi\mathcal{N}}\rho e^{i\pi\mathcal{N}}\Tilde{\eta}_{k}^{\dagger}-\frac{1}{2}\left\{\Tilde{\eta}_{k}^{\dagger}\Tilde{\eta},\rho\right\}\right]\\
        \\
        &+B_{R}^{k}\left[\Tilde{\eta}_{k}^{\dagger}e^{i\pi\mathcal{N}}\rho e^{i\pi\mathcal{N}}\Tilde{\eta}_{k}-\frac{1}{2}\left\{\Tilde{\eta}_{k}\Tilde{\eta}_{k}^{\dagger},\rho\right\}\right]
    \end{aligned}\quad,
\end{equation*}

where we define

\begin{equation}\label{eq:15}
    \begin{aligned}
        A_{L}^{k}=\gamma\left(g_{Lk}\right)^{2}\chi_{L,k}(1-f_{L,k})\quad& B_{L}^{k}=\gamma\left(g_{Lk}\right)^{2}\chi_{L,k}f_{L,k}\\
        \\
        A_{R}^{k}=\gamma\left(g_{Rk}\right)^{2}\chi_{R,k}(1-f_{R,k})\quad& B_{L}^{k}=\gamma\left(g_{Rk}\right)^{2}\chi_{R,k}f_{R,k}\quad.
    \end{aligned}
\end{equation}

Our system is  driven out of equilibrium by thermal reservoirs and every site subjected to dephasing noise. The time evolution of the density matrix is described via a Lindblad master equation with the new depahsing dissipator

\begin{equation}\label{eq:16}
    \mathcal{D}_{i}^{deph}(\rho)=\frac{\Gamma}{2}\left(\sigma_{i}^{z}\rho\sigma_{i}^{z}-\rho\right)
\end{equation}
where $\Gamma$ is the dephasing strength.

In the presence of dephasing, the Lindblad equation is given by the following representation

\begin{equation}\label{eq:17}
    \frac{d}{dt}\rho=-i\left[H,\rho\right]+\sum_{r=L,R}\mathcal{D}_{r}(\rho)+\sum_{i} \mathcal{D}_{i}^{deph}(\rho)\quad,
\end{equation}
where $\mathcal{D}_{r}(\rho)$ are the dissipators given by Eq. \eqref{eq:12} and Eq. \eqref{eq:13}.

\subsection{Non Equilibrium Steady-state equation for the Covariance Matrix}

The fermionic nature of the model allows us to focus on the steady state properties only on the system's covariance matrix defined as

\begin{equation}\label{eq:18}
    C_{ij}=\left<\eta_{j}^{\dagger}\eta_{i}\right>\quad,
\end{equation}

From Eq. \eqref{eq:17} we can write the following time evolution for the operator $\left<\eta_{n}^{\dagger}\eta_{m}\right>$:

\begin{equation}\label{eq:19}
    \begin{aligned}
        \frac{d}{dt}\left<\eta_{n}^{\dagger}\eta_{m}\right>=&i\left<\left[\mathcal{H},\eta_{n}^{\dagger}\eta_{m}\right]\right>+Tr\left\{\eta_{n}^{\dagger}\eta_{m}\mathcal{D}^{L}\right\}+\\
        \\
        &+Tr\left\{\eta_{n}^{\dagger}\eta_{m}\mathcal{D}^{R}\right\}+\sum_{i}Tr\left\{\eta_{n}^{\dagger}\eta_{m}\mathcal{D}_{i}^{deph}\right\}\quad.
    \end{aligned}\quad.
\end{equation}

Note that the covariance matrix in Eq. \eqref{eq:18} is given in terms of the fermionic operators $\eta$ $(\eta^{\dagger})$ and the dissipators in Eq. \eqref{eq:12} and \eqref{eq:13} are given in terms of the new set of fermionic operators $\Tilde{\eta}$ $(\Tilde{\eta}^{\dagger})$.

The time evolution for the covatiance matrix is 

\begin{equation}\label{eq:20}
     \frac{d}{dt}C=-\left[WC+CW^{\dagger}\right]+F-\Gamma\Delta(C)\quad,
\end{equation}
where $W$ and $F$ are a temperature dependent matrix given by $ W=iH+S^{-1}\mathcal{M}S$:

\begin{equation*}
    \mathcal{M}=\frac{1}{2}diag\left(A_{1}^{L}+A_{1}^{R}+B_{1}^{L}+B_{1}^{R},...\right)
\end{equation*}

\begin{equation*}
    \mathcal{B}=diag\left(B_{1}^{L}+B_{1}^{R},B_{2}^{L}+B_{2}^{R},...\right)\quad
\end{equation*}

\begin{equation*}
    F=S\mathcal{B}S^{-1}\quad,
\end{equation*}

In Eq. (20),  $\Delta(C)$ is an operation that removes the diagonal elements of a matrix:

\begin{equation}\label{eq:21}
    \Delta(C)=C-diag(C_{11},C_{22},\dots,C_{NN})\quad.
\end{equation}

In the \textit{NESS}, dC/dt=0, which give us the matrix equation

\begin{equation}\label{eq:22}
    WC+CW^{\dagger}+\Gamma\Delta(C)=F\quad.
\end{equation}
\noindent

Note that, when $\Gamma=0$, this reduces to a Lyapunov equation

\begin{equation}\label{eq:23}
    WC+CW^{\dagger}=F\quad.
\end{equation}

Due to the nature of matrix $\mathcal{M}$ and $\mathcal{B}$ we are able to solve systems up to N = 100. When $\Gamma\ne 0$,
Eq. (21) is still linear in C, but not in Lyapunov-form and the complexity of the matrix remain the same (we need the eigenvalues and eigenvector of $H$ to solve the system).

\section{Transport properties with dephasing}

The classification of the transport regime can be caracterized, in general, as a power-law scalling with the system size:

\begin{equation}\label{eq:24}
    J\varpropto\frac{1}{N^{\alpha}}\quad,
\end{equation}
where $\alpha\ge0$ is a transport coefficient. The transport is classified as ballistic, diffusive and anomalous wich corresponds to $\alpha=0$, $\alpha=1$ and $\alpha>1$ or $\alpha<1$ respectively. In the literature we have many works devoted to study the behavior of the current in the presence of  dephasing bulk in \textit{Boundary-Driven} systems. For a recent review, see \cite{gtlandi}.

As an exemple, we cite Ref. \cite{zn,asadian} which showed the change of the transport regime in the termodinamic limit with any non-zero dephasing strengths and Ref. \cite{arthur} which showed that quasi-periodic potentials  rise an enhancement of transport, increasing the systems conductivity.

In this context, our focus here is to go beyond \textit{Boundary-Driven} systems and study the behavior of the spin current in the presence of dephasing and thermal baths to induce the system out of the equilibrium.

\subsection{Spin/Particle Current}

Using Eq. \eqref{eq:19} we can derive an expression for the particle currents. In the fermionic representation, the temperature imbalance between the two baths will lead to a flow of particles along the chain. In the spin representation, this is mapped into a flow of magnetization. 

To evaluate the current of particles/magnetization, we start with a conservation law for the time evolution of $\left<\mathcal{N}\right>$, where $\mathcal{N}$ is the total number of fermions:

\begin{equation*}
    \mathcal{N}=\sum_{i=1}^{N}\Tilde{\eta}_{i}^{\dagger}\Tilde{\eta}_{i}.
\end{equation*}

Since $[H,\mathcal{N}] = 0$, it follows from Eq. \eqref{eq:19} that

\begin{equation}\label{eq:25}
\begin{aligned}
    \frac{d}{dt}\left<\mathcal{N}\right>=\frac{d}{dt}Tr\left\{\mathcal{N}\rho\right\}=&Tr\left\{\mathcal{N}\mathcal{D}_{L}(\rho)\right\}+Tr\left\{\mathcal{N}\mathcal{D}_{R}(\rho)\right\}\\
    \\
    &+Tr\left\{\mathcal{N}\mathcal{D}^{deph}(\rho)\right\}\quad.
    \end{aligned}
\end{equation}

Note that, the dephasing dissipator is given by the fermionic operators, then using the relation between $\eta$ and $\Tilde{\eta}$ \eqref{eq:10}, we can compute the contribution of  $Tr\left\{\mathcal{N}D_{i}^{deph}\right\}$. Using some properties of the fermionic operators it is easy to show that $Tr\left\{\mathcal{N}D_{i}^{deph}\right\}=0$, then the dephasing term does not affect the continuity equation. Hence, in \textit{NESS} we have

\begin{equation}\label{eq:26}
    \begin{aligned}
            J_{\mathcal{N}}&=Tr\left\{\mathcal{N}\mathcal{D}_{L}(\rho)\right\}=-Tr\left\{\mathcal{N}\mathcal{D}_{R}(\rho)\right\}\\
            \\
            &=\sum_{k}\gamma g_{L,k}\chi_{L,k}\left[f_{L,k}-\left<\Tilde{\eta}_{k}^{\dagger}\Tilde{\eta}_{k}\right>\right]\quad.
    \end{aligned}
\end{equation}

Defining $\Tilde{C}_{ij}=\left<\Tilde{\eta}_{j}^{\dagger}\Tilde{\eta}_{i}\right>$ and using Eq. \eqref{eq:10} we find

\begin{equation}\label{eq:27}
    \begin{aligned}
        \Tilde{C}_{ij}&=\left<\Tilde{\eta}_{j}^{\dagger}\Tilde{\eta}_{i}\right>\\
        \\
             &=\sum_{kl}S_{ik}C_{kl}S_{lj}^{-1}\\
        \\
              &=\left(SCS^{-1}\right)_{ij}\quad,\\
         \\
    \end{aligned}
\end{equation}
then the covariance matrix in the momemtum representation is given by $\Tilde{C}=SCS^{-1}$, where $S$ is given in terms of the eigenvectors of H. With this result, we can write the particule current as

\begin{equation}\label{eq:28}
     J_{\mathcal{N}}=\sum_{k}\gamma g_{L,k}\chi_{L,k}\left[f_{L,k}-\Tilde{C}_{k,k}\right]\quad.
\end{equation}

This expression is more general and we can use it even when including the dephasing pertubation in Eq. \eqref{eq:22}.
\section{Numerical Analysis}
\subsection{Spin current in Homogeneous chain}

First of all to better understand the interplay between dephasing and the strucuture of the Hamiltonian we study the behavior of a homogenous chain subject to dephasing noise. The Hamiltonian is described by the follwing interactions

  \begin{equation}\label{eq:29}
  \begin{aligned}
W_{ii}&=2\\
\\
W_{i+1,i}&=W_{i,i+1}=1
  \end{aligned}
    \end{equation}
\noindent
that is, we have a homogeneous magnetic field $h_{i}=2$ and homogeneous interactions $\alpha_{i}=1$. The spin current is depicted in Fig. 1.

\begin{figure}[ht!]
    \centering
    \includegraphics[scale=0.65]{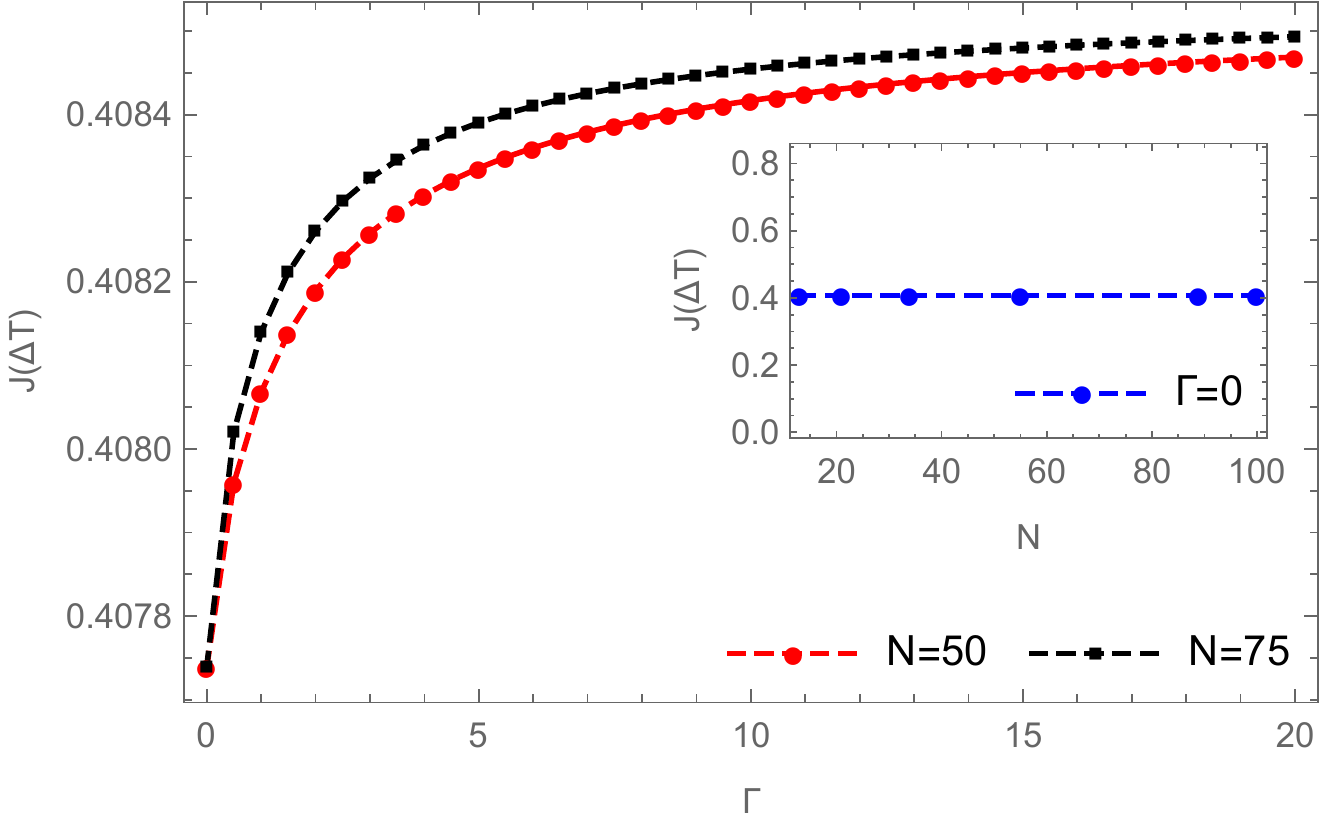}
    \caption{Spin Current as a function of the dephasing strength $\Gamma$ for differents values of N. The temperature gradient is fixed at $\Delta T=45$. We see that the forward current have a small enhancement (less than $1\%$) even in strong dephsaing rate.}
    \label{fig:1}
\end{figure}

When dephasing is present, we have a small enhancement of current (less than $1\%$) and the balistc behavior is preserved. As we will see in the next section the interplay between the graded structure of the Hamiltonian and the dephasing strengh rises non trivial phenomenom.

\subsection{Dephasing Enhanced Rectification}

Now we consider the first case of our numerical analysis. The Hamiltoninan is graded and described by:

\begin{equation}\label{eq:30}
    h_{i}=1\quad\quad\quad  \alpha_{i}=N-\left(\frac{i-1}{N-1}\right)N+1\quad,
\end{equation}
that is, the magnetic field is homogeneous and the spin interaction is decreasing linearly from site 1 to N. We can write

\begin{equation}\label{eq:31}
\begin{aligned}
    H=\sum_{n,m}W_{n,m}\eta_{n}^{\dagger}\eta_{m}\quad,
    \end{aligned}
\end{equation}
with $W_{n,m}$ given by entries $W_{j,j}=1$ and $W_{j,j+1}=W_{j+1,j}=\alpha_{j}$.

For the system to present the rectification phenomenon, it is essential that it presents asymmetry in the interaction between the spins. As soon as we invert the thermal baths, we see different flows. As we discussed in the introduction, it is possible to build devices where we can  efficiently handle  the magnitude of the current. Specifically, if the left-right symmetry is broken, the magnitude of the spin current, which is  given by $J(\Delta T)$ and induced by a positive bias, may be different with respect to the magnitude of $J(-\Delta T)$. We define the rectification factor R as
\begin{equation}\label{eq:32}
    R=\frac{J(\Delta T)+J(-\Delta T)}{J(\Delta T)-J(-\Delta T)}
\end{equation}
for $\Delta T>0$ we have $J(\Delta T)>0$ (then $J(-\Delta T)<0$). This definition is important to see the efficiency  of rectification. $R=0$ means that no rectification takes place ($J(\Delta T)=J(-\Delta T))$, while $\left|R\right| = 1$ means that we have perfect rectification (i.e. the current is finite in one direction, and null in the other). Other values of R, positive (negative) indicate that the flow is greater for positive (negative) temperature biases.

According to Eq. \eqref{eq:28} we find the pattern for the spin current shown in Fig. 2:

\begin{figure}[ht!]
    \centering
    \includegraphics[scale=0.7]{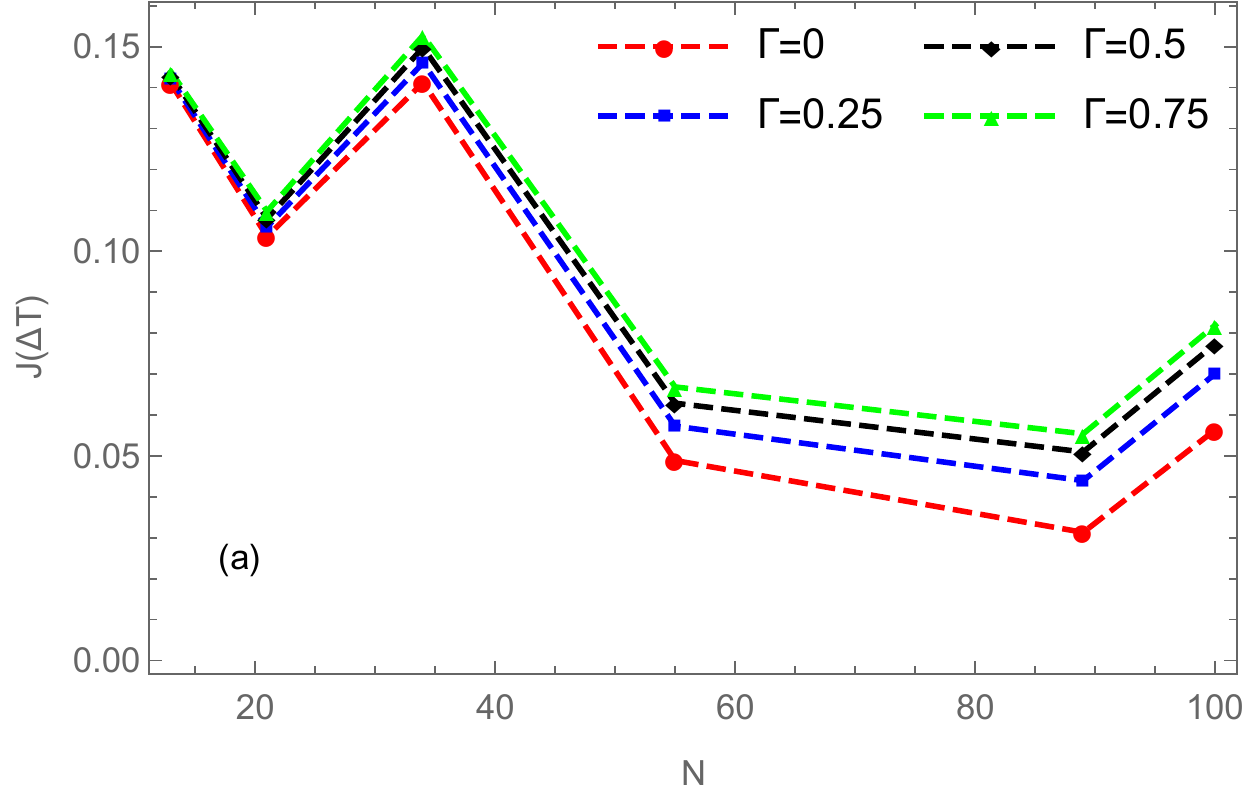}
    \label{fig:my_label}
\end{figure}
\begin{figure}[ht!]
    \centering
    \includegraphics[scale=0.7]{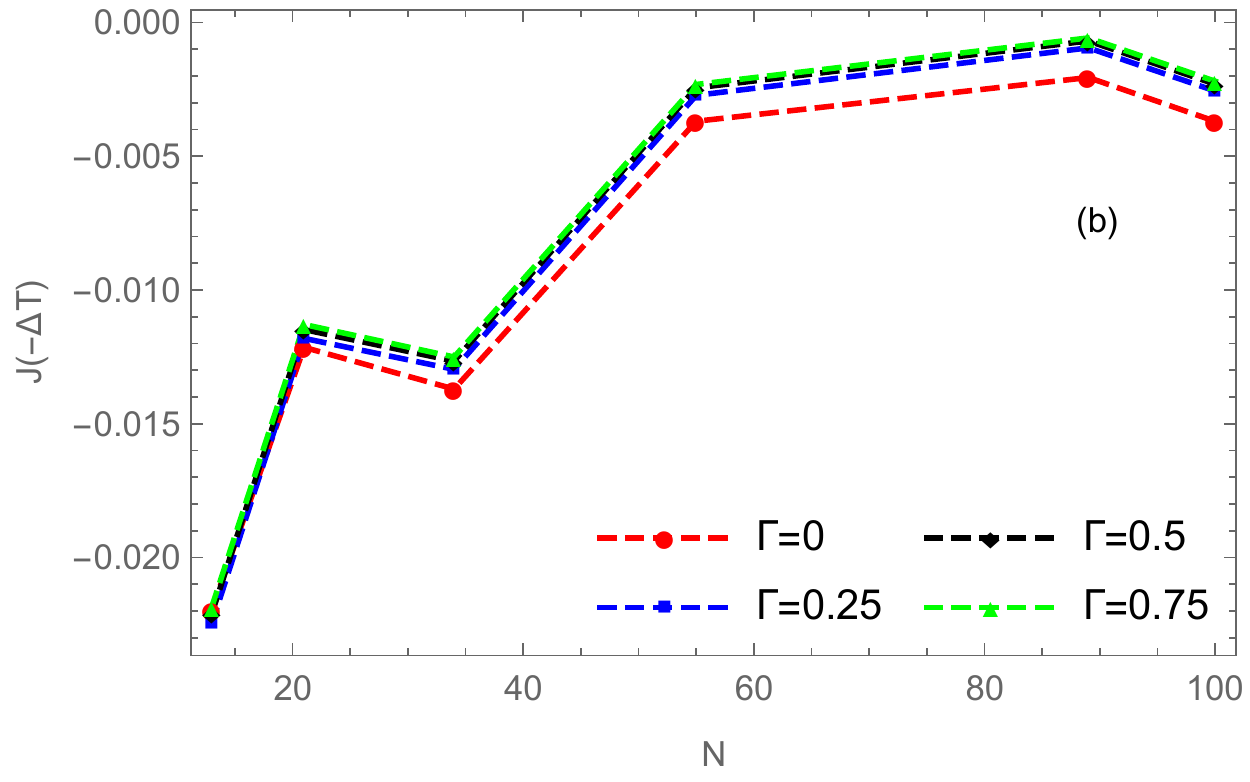}
    \caption{Spin Current for differents values of dephasing. The temperature gradient is fixed at $\Delta T=95$. We see that the forward current is enhanced and the backward current is erased when we grow the dephasing rate.}
    \label{fig:2}
\end{figure}

We see that when dephasing is present the forward current is enhanced (Fig.2 (a)) and the backward current ($\Delta T<0$)) is supressed (Fig. 2 (b)), then we get a rectification enhanced phenomenom depicted in Fig. 3

\begin{figure}[ht!]
    \centering
    \includegraphics[scale=0.7]{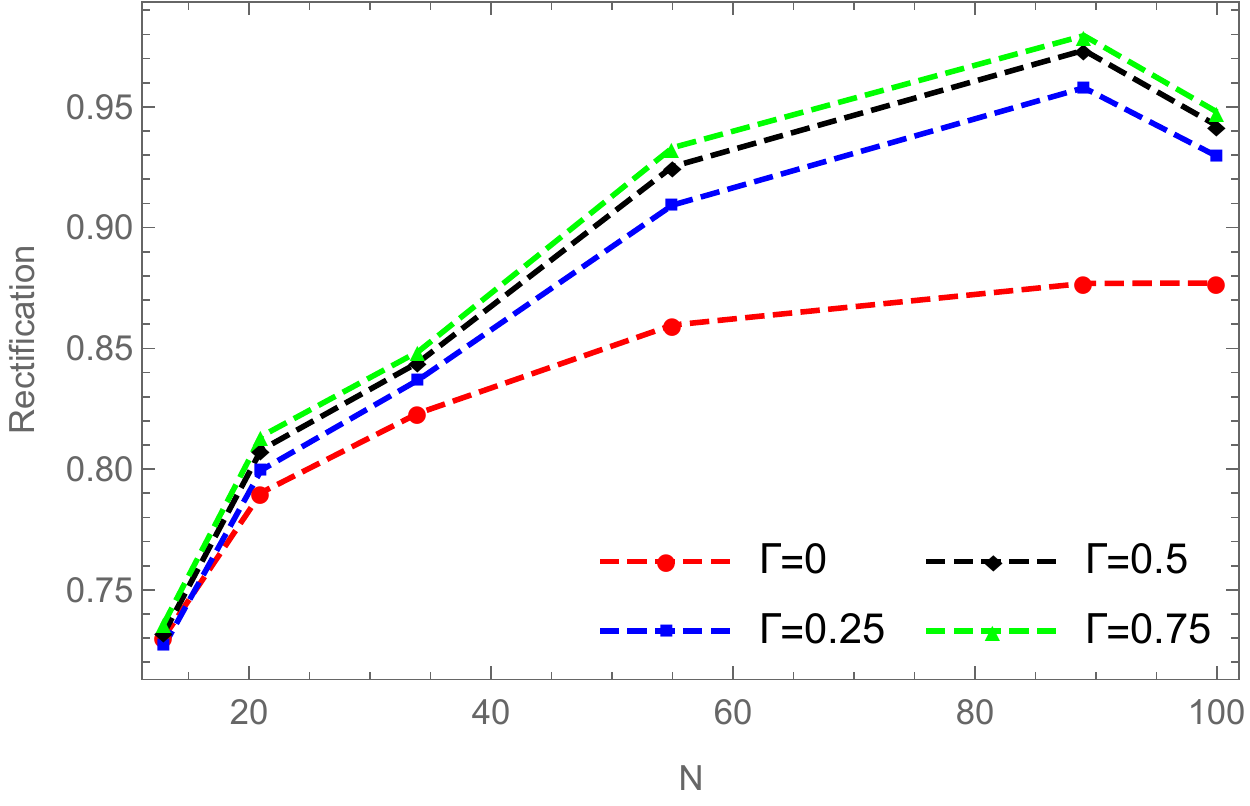}
    \caption{Rectification pattern of a graded chain described by Eq. \eqref{eq:30} for differents values of dephasing. The temperature gradient is set at $\Delta T=95$. When we let the dephaisng rate assume more intense values, the rectification is enhanced.}
    \label{fig:3}
\end{figure}

We also study the behavior of currents in the presence of strong magnetic field, that is

\begin{equation}\label{eq:33}
    h_{i}=10\quad\quad \alpha_{i}=N-\left(\frac{i-1}{N-1}\right)N\quad,
\end{equation}

\begin{figure}[ht!]
    \centering
    \includegraphics[scale=0.7]{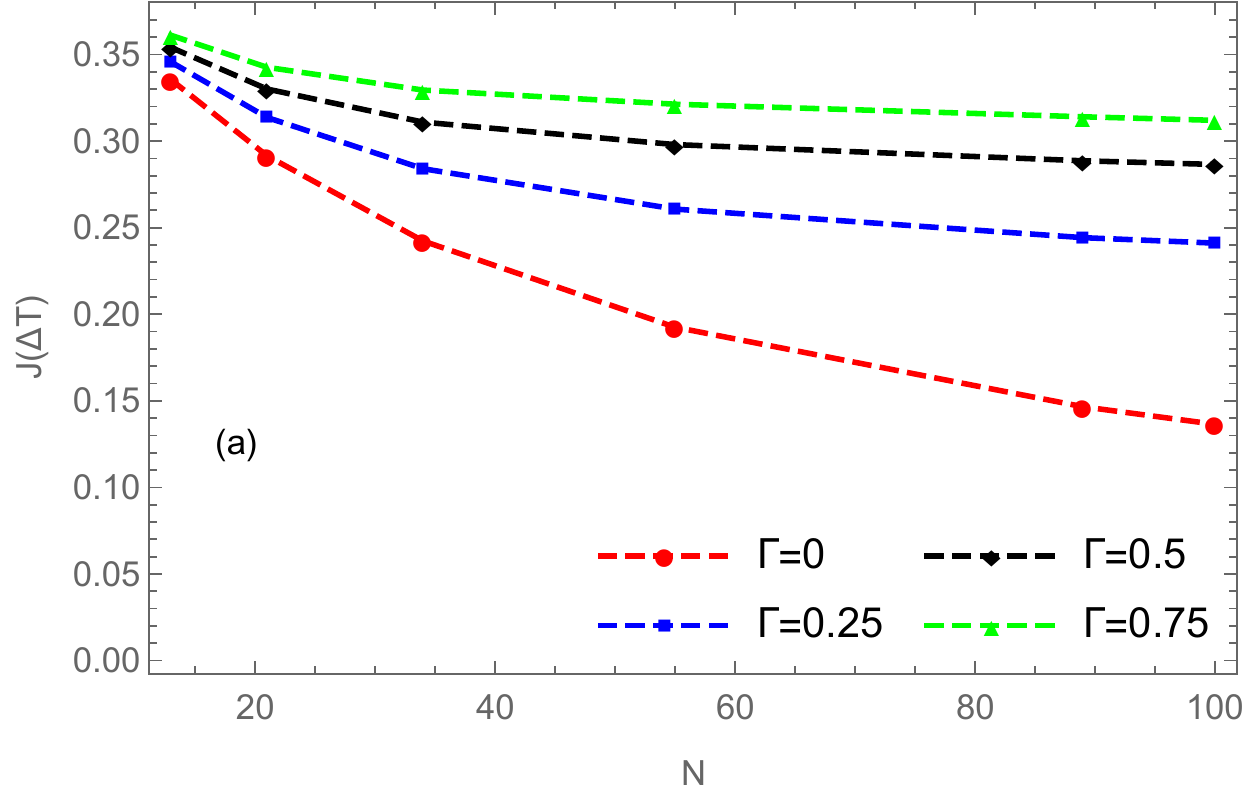}
\end{figure}
\begin{figure}[ht!]
    \centering
    \includegraphics[scale=0.7]{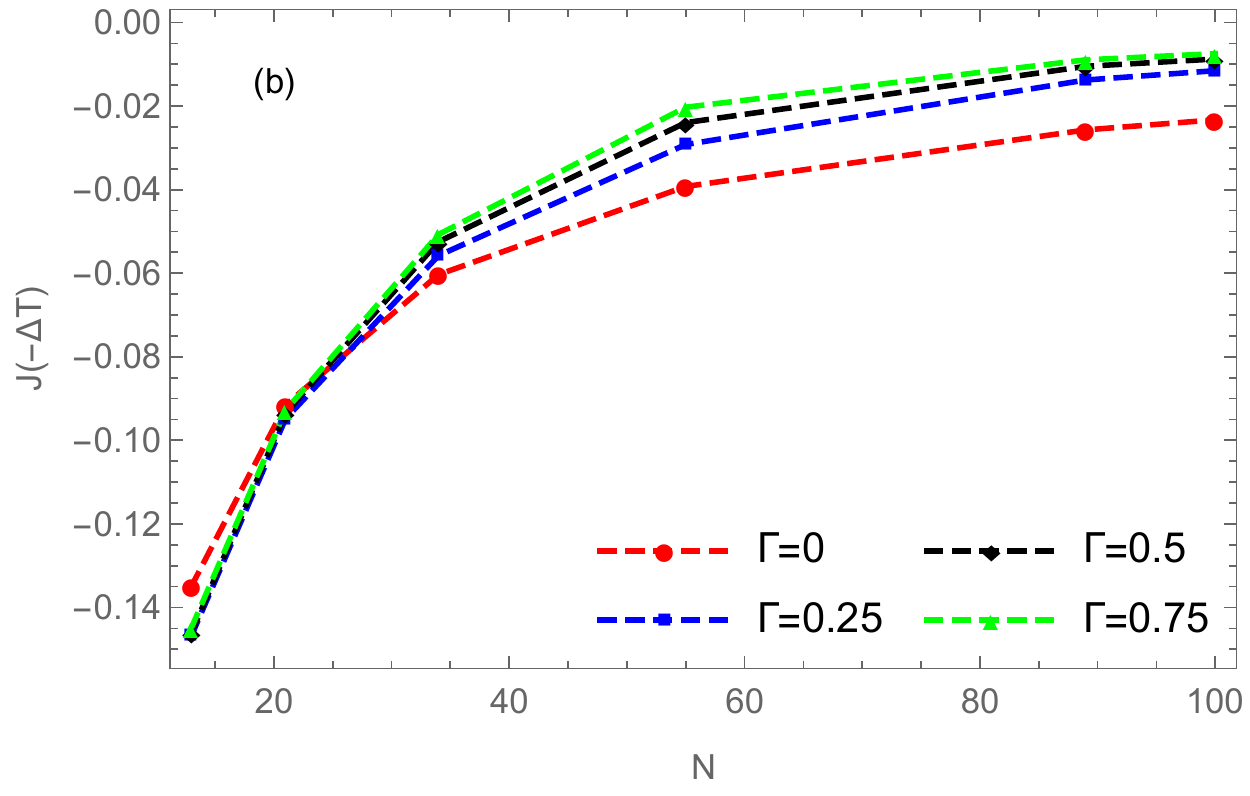}
    \caption{Spin Current for differents values of dephasing with intense magnetic field. The temperature gradient is fixed at $\Delta T=95$. Again we see that the forward current is enhanced and the backward current is erased when we grow the dephasing rate.}
    \label{fig:4}
\end{figure}

\begin{figure}[ht!]
    \centering
    \includegraphics[scale=0.7]{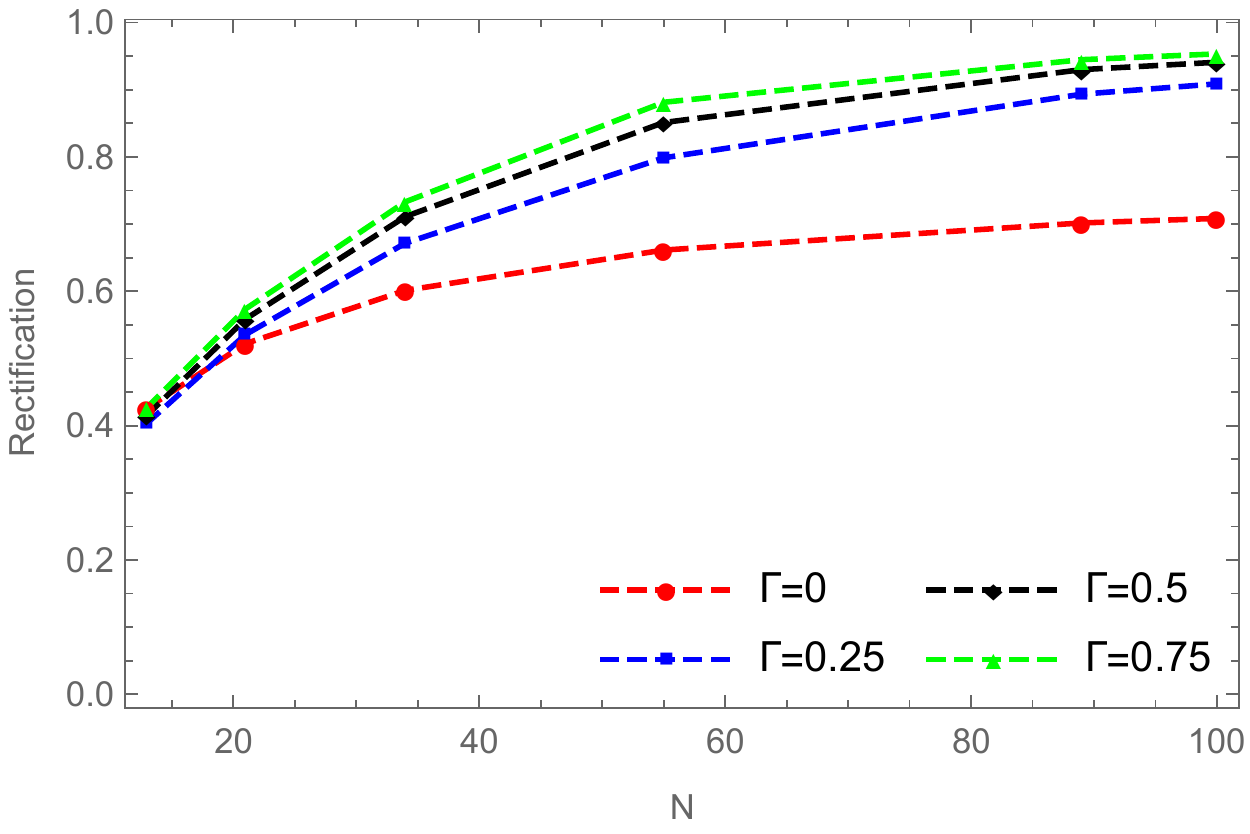}
    \caption{Rectification pattern of a graded chain described by Eq. \eqref{eq:33} for differents values of dephasing. The temperature gradient is set at $\Delta T=95$. When we let the dephaisng rate assume more intense values, the rectification is enhanced. for strong magnetic field.}
    \label{fig:5}
\end{figure}

Again we find that, when dephasing is present, the forward current is enhanced  and the backward current ($\Delta T<0$)) is erased (Fig. 4 (b)). Then we get a rectification enhanced phenomenom assuming a value closer to one when the magnetic field is stronger, Fig. 5.
\newpage
\subsection{One Way Street of Spin Current}

Now we focus on a fully graded system given by

\begin{equation}\label{eq:34}
    h_{i}=1+\left(\frac{i-1}{N-1}\right)N\quad\quad  \alpha_{i}=N-\left(\frac{i-1}{N-1}\right)N\quad,
\end{equation}
that is, the Hamiltonian is composed by a graded magnetic field growing lineraly from site 1 to $N$, and with spin interaction decreasing lineary from site 1 to N. In this situation we introduce more assimetry in the system, then the interplay between the Hamiltonian and the dephasing term becomes more complex to analyze, that is, with a graded system the coherences are affected. 

Using the expression for the spin current in Eq.\eqref{eq:28} we find the pattern depicted in Fig. 6 and 7. We see that when dephasing is present, the inverse spin current (when the temperature gradient is inverted) assume a positive value. This phenomenom is dependent on three parameters: the size of the chain (N), the intensity of the dephasing rate ($\Gamma$) and the temperature gradient ($\Delta T$). 

It is interesting to mention that the one way phenomenon has already been observed in other systems. As an example, we cite the one way street phenomenon for the energy current in the $XXZ$ model submitted to spin reservoirs at the ends\cite{oneway}.
\begin{figure}[ht!]
    \centering
    \includegraphics[scale=0.57]{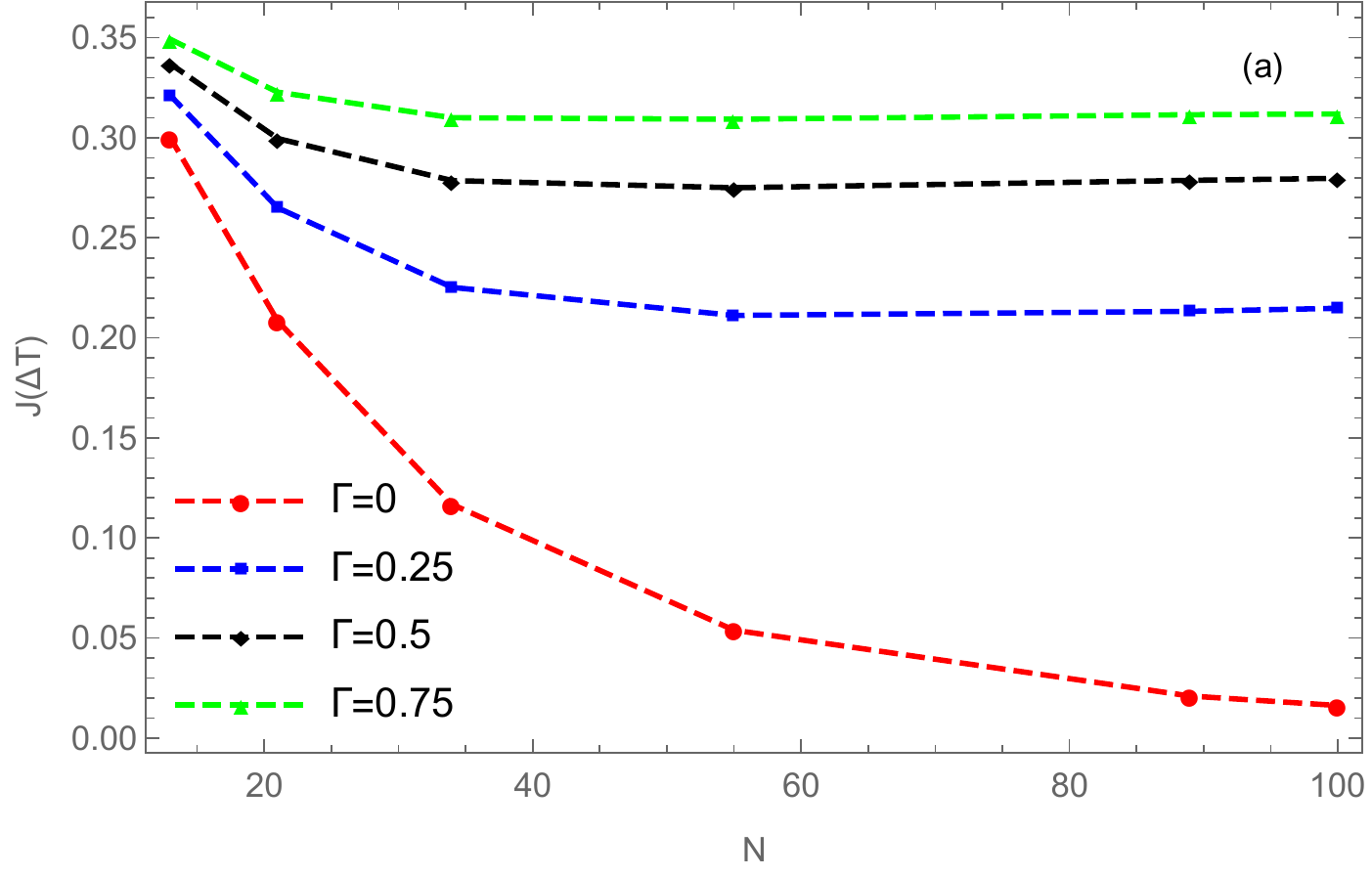}
\end{figure}

\begin{figure}[ht!]
    \centering
    \includegraphics[scale=0.65]{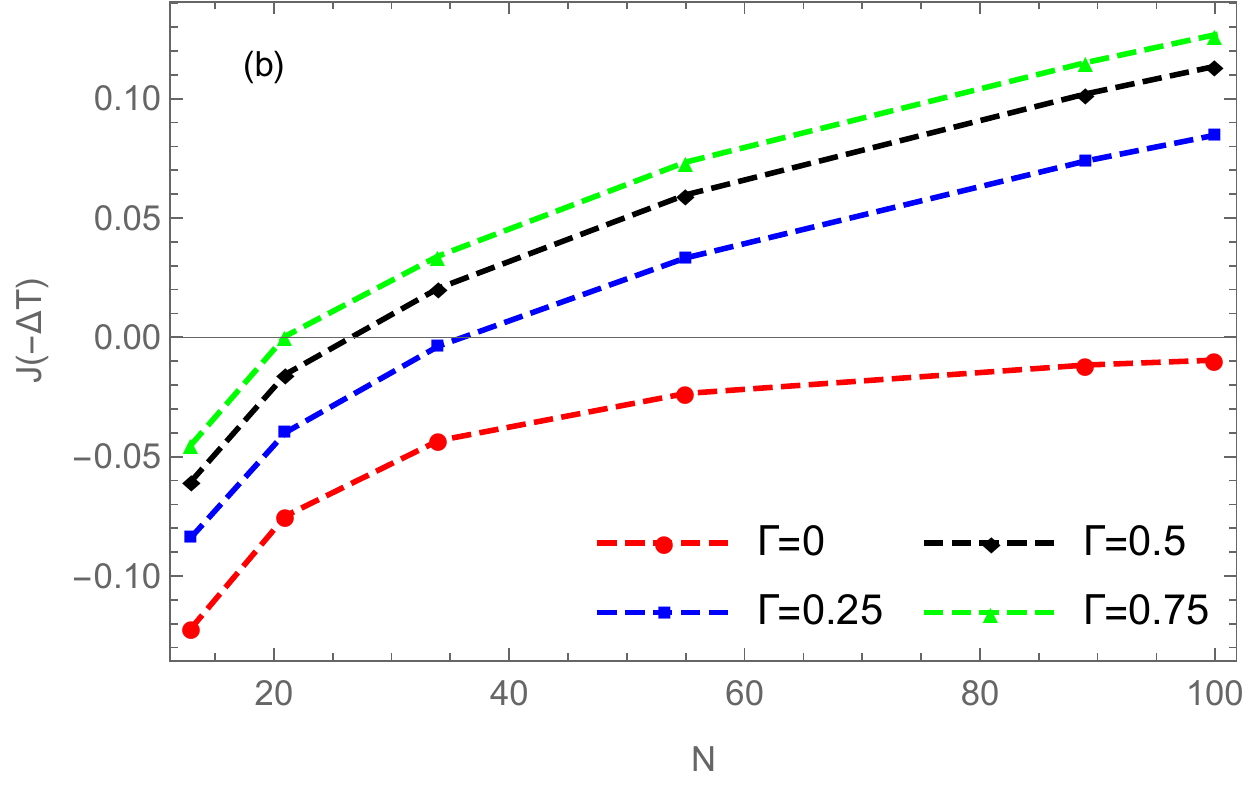}
    \caption{Spin Current for differents values of dephasing. The temperature gradient is set at $\Delta T=95$. The one way phenomenom occurs depending on $N$, $\Delta T$ and $\Gamma$}
    
    \label{fig:6}
\end{figure}

\begin{figure}[ht!]
    \centering
    \includegraphics[scale=0.65]{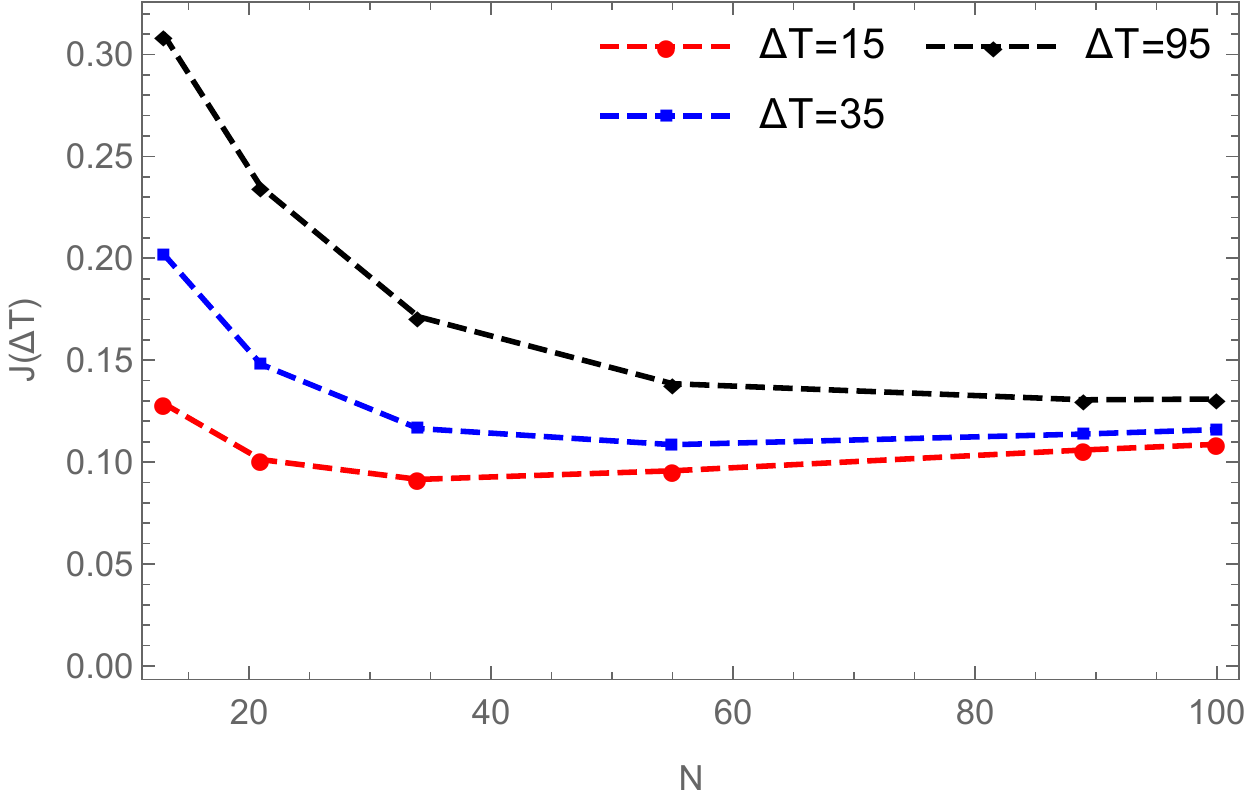}
        \caption{Forward spin Current for different values of temperature. The dephasing is set at $\Gamma=0.1$.}
    \label{fig:07}

\end{figure}
\begin{figure}[ht!]
    \centering
    \includegraphics[scale=0.65]{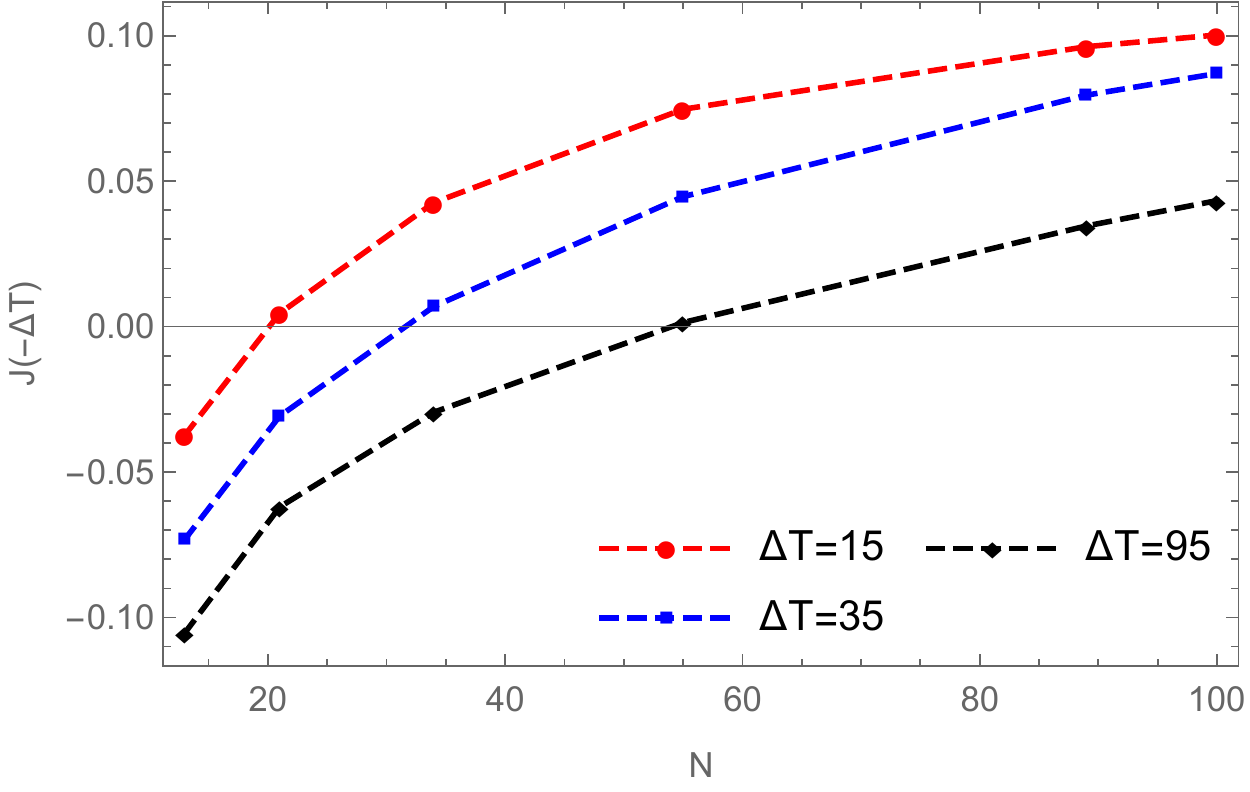}
    \caption{Backward spin Current for different values of temperature. The dephasing is set at $\Gamma=0.1$. We see that with a specif choice of $N$, $\Gamma$ and $\Delta T$ it is possible to get almost perfect rectification, that is, the inverse current $J(-\Delta T)$ is going to zero}
    \label{fig:7}
\end{figure}

\newpage
\section{Final Remarks}

In the present paper, aiming to understand effective mechanisms to manipulate and control currents in quantum systems, we investigate in detail the spin current in the $XX$ chain subject to graded and nearest neighbor interactions and  global dissipators. When the system is subject to dephasing noise, we show the existence of non trivial behavior of spin current, that is, we show the existence of rectification enhancement mechanims and how it is possible to control the spin currrent through internal parameters of the microscopic evolution that the system is subject ($\Delta T$, $N$ and $\Gamma$). We also observe that it is possible to obtain perfect rectification using graded materials and specific choice of parameters (see  Fig. 6 and Fig. 8).

It is also worth to recall that the $XX$ chain has already shown to be an effective system to obtain rectification \cite{sxx}. It is interesting to comment that these microscopic systems can be performed experimentally, as an example we can cite a more complex system, the $XXZ$ chain, with different  values for the coefficientes of $\sigma_{j}^{x}\sigma_{j+1}^{x}, \sigma_{j}^{y}\sigma_{j+1}^{y}$ and $\sigma_{j}^{z}\sigma_{j+1}^{z}$ \cite{endres, barredo}. 

Another relevant comment is that the Heisenberg model, which has central role in analytical results, can be experimentally simulated (implemented). As an example we mention the study of energy transport by means of cold atoms in optical lattices \cite{bloch} or trapped ions \cite{blatt}. And experiments with Rydberg atoms in optical traps involving these spin models are presented  in Ref.\cite{duan, whitlock, PhysRevX}.

To conclude, with the results presented here we believe to present some light in the understanding of the dephasing effects in the currents of spin chains, and these results will be certainly useful in the problem of manipulation of the currents. Moreover, we believe that the occurrence of  rectification in this simple model indicates an ubiquitous phenomenon in spin systems.  

\textbf{Acknowledgment:} Work partially supported by
CNPq (Brazil).


\end{document}